\documentstyle[11pt,aaspp4]{article}

\begin{document} 

\hyphenation{PSPC}
\hyphenation{X-ray}

\slugcomment{To Appear in ApJ September 10, 1996}

\title{The Twisting X-ray Isophotes of the Elliptical Galaxy NGC 720} 

\author{David A. Buote\altaffilmark{1} and Claude R. Canizares}

\affil{Department of Physics and Center for Space Research 37-241, \\
Massachusetts Institute of Technology \\ 77 Massachusetts Avenue,
Cambridge, MA 02139, \\ dbuote@space.mit.edu, crc@space.mit.edu}

\altaffiltext{1}{Present Address: Institute of Astronomy, Madingly
Road, Cambridge CB3 0HA, UK}

\begin{abstract}

We present spatial analysis of the deep (57ks) ROSAT HRI X-ray image
of the E4 galaxy NGC 720. The orientation of the HRI surface
brightness is consistent with the optical position angle $(PA)$
interior to semi-major axis $a\sim 60\arcsec$ (optical $R_e\sim
50\arcsec$). For larger $a$ the isophotes twist and eventually
$(a\gtrsim 100\arcsec)$ orient along a direction consistent with the
$PA$ measured with the PSPC data (Buote \& Canizares 1994) -- the
$\sim 30\arcdeg$ twist is significant at an estimated $99\%$
confidence level. We argue that this twist is not the result of
projected foreground and background sources, ram pressure effects, or
tidal distortions.  If spheroidal symmetry and a nearly isothermal hot
gas are assumed, then the azimuthally averaged radial profile displays
features which, when combined with the observed $PA$ twist, are
inconsistent with the simple assumptions that the X-ray emission is
due either entirely to hot gas or to the combined emission from hot
gas and discrete sources.  We discuss possible origins of the $PA$
twist and radial profile features (e.g., triaxiality).

\end{abstract}

\section{Introduction\label{intro}}

The intrinsic shapes of elliptical galaxies are currently not well
understood (e.g., de Zeeuw \& Franx 1991; de Zeeuw 1995). Hydrodynamic
simulations of galaxy halos in a Cold Dark Matter universe typically
produce flattened $(\epsilon\sim 0.5)$, oblate-triaxial halos (e.g.,
Katz \& Gunn 1991; Dubinski 1994). Yet there is a paucity of
observational constraints for the intrinsic shapes and, in particular,
the degree of triaxiality in ellipticals. For a handful of galaxies
the intrinsic shapes have been precisely measured and suggest highly
elongated halos with $\epsilon\sim 0.6$ (e.g., Sackett 1995), though
the degree of triaxiality has not been constrained.  Recently, Statler
(1994a,b) has shown that in principle the stellar velocity fields can
effectively probe triaxiality of ellipticals, however the observations
required are extensive and the current constraints allow for a large
range of triaxial shapes.

The E4 galaxy NGC 720 is a promising candidate for analysis of the
shapes of its X-ray isophotes because it should not be substantially
affected by ram pressure or tidal distortions due to it being very
isolated from other large galaxies (Schechter 1987). Moreover, NGC 720
is quite bright in the ROSAT band ($8\times 10^{-13}$ erg cm$^{-2}$
s$^{-1}$ for 0.5-2.0 keV) and is likely to be dominated by emission
from hot gas as indicated by its large ratio of X-ray to optical
luminosity (Kim, Fabbiano, \& Trinchieri 1992).  Also, radio emission
has not been detected from NGC 720 (e.g., Birkinshaw \& Davies 1985)
indicating that the gas should not be substantially disturbed by
magnetic fields.  Since environmental effects should be unimportant,
the X-ray--emissivity should trace the shape of the three-dimensional
gravitational potential, independent of the temperature profile of the
gas (Buote \& Canizares 1994, 1996); though possible rotation of the
hot gas may affect the emissivity shape in the core (see \S
\ref{anal}).  Thus, the X-ray isophote shapes and orientations probe
the intrinsic shape of the potential, and hence the mass, of the
galaxy.

From analysis of ROSAT Position Sensitive Proportional Counter (PSPC)
data of the E4 galaxy NGC 720 Buote \& Canizares (1994, hereafter BC)
concluded that the isophote shapes and radial profile of the X-ray
emission requires an extended, massive, flattened dark halo
$(\epsilon\sim 0.5-0.7)$. However, the X-ray isophotes at $a\sim
100\arcsec$ are misaligned by $\sim 30\arcdeg$ from the optical
isophotes, a fact that was not incorporated into the analysis of BC.
The orientations of the inner X-ray isophotes $(a\lesssim 60\arcsec)$
are not well determined by the PSPC data because of the width of the
point spread function (PSF) ($\sim 30\arcsec$ FWHM). To better
understand the nature of this measured misalignment between X-ray and
optical isophotes, we obtained a high resolution observation to
specifically address the orientation of the X-ray contours at small
radii.

\section{Observations and Data Analysis \label{obs}}

From January 8-12, 1994 NGC 720 was observed with the High Resolution
Imager (HRI) on board ROSAT (Tr\"{u}mper 1983) resulting in an
effective exposure of 57ks; for descriptions of the ROSAT X-ray
telescope see Aschenbach (1988) and the HRI see David et
al. (1995). The spatial resolution of the HRI is $\sim 4\arcsec$ FWHM,
but the HRI possesses only limited spectral resolution in its 0.1 -
2.4 keV pass-band.

To prepare the HRI image for spatial analysis we (1) rebinned the
image into a more manageable $1638\times 1638$ field of $2.5\arcsec$
pixels, (2) searched for time intervals where the background was
anomalously high, (3) flattened the image, (4) selected PHA channels
1-5 to optimize signal-to-noise ratio ($S/N$), (5) examined the
accuracy of the aspect solution, (6) removed foreground and background
sources embedded in the galactic continuum, and (7) subtracted the
background. All of the reduction procedures were implemented with the
standard IRAF-PROS and FTOOLS software.

By masking out detected sources in the HRI field (including NGC 720
itself using a circle of $10\arcmin$ radius), we constructed the
background light curve which shows no significant fluctuations above
those expected for the HRI due to external radiation (David et
al. 1995). The image was flattened with the exposure map generated by
the task {\it hriexpmap} in FTOOLS; note that the true resolution of
this exposure map actually corresponds to $5\arcsec$ pixels.  The
ground calibration results of David et al. (1995) show that several
PHA channels have low $S/N$. We examined the effects of the PHA
channels on $S/N$ for a circle of $50\arcsec$ radius centered on NGC
720; the background was taken from an annulus centered on NGC 720 with
$(r\sim 7\arcmin, \Delta r = 1\arcmin)$ with sources masked
out. Including only PHA channels 1-5 optimized the $S/N$ which is
consistent with the results of David et al., with the possible
exception of channel 1 for which David et al. concluded also has low
$S/N$.

Since we are concerned in this paper with measurements of position
angles and ellipticities of the X-ray isophotes in NGC 720 an accurate
aspect solution for the HRI observation is essential.  The point
sources in the field are not sufficiently bright to usefully perform
the aspect correction algorithm proposed by Morse (1994). Hence, we
examined the aspect solution individually for the different
observational intervals (OBIs); i.e. those time intervals where the
spacecraft continuously pointed on NGC 720.  From analysis of the
point sources in the fields of the three OBIs with appreciable
exposures we find no statistically significant relative shifts of the
OBIs. Typically, the required shifts were $0.4 \pm 0.8$ pixels
(i.e. $1\arcsec \pm 2\arcsec$).  The effects of the small shifts
consistent with this result are insignificant when considering the
measurement of ellipticities and position angles on scales of interest
to this study (i.e. $\gtrsim 15\arcsec$).

There is one point source $150\arcsec$ to the SE, another $200\arcsec$
to the NW, and 4 more straddling the circle of $400\arcsec$ radius
from the galaxy; within $\sim 2\arcmin$ of the center of NGC 720 there
are no obvious point sources.  Since we cannot obtain useful
constraints on the ellipticity and position angle of the X-ray
isophotes for $r>100\arcsec$ due to inadequate $S/N$ we simply remove
these sources. We removed sources by first choosing an annulus around
each source to estimate the local background. We then fitted a second
order polynomial surface to the background and replaced the source
with the background. This procedure is well suited for estimating
quadrupole moments of high $S/N$ cluster images; see Buote \& Tsai
(1996) for a thorough discussion. The effects of unresolved points
sources on the shapes and orientations of the isophotes are examined
in the next section.

\begin{figure}[p]
\caption{  \label{fig.image} }
\raggedright

Contour map of the HRI X-ray surface brightness of the E4 galaxy
NGC 720 binned into $5\arcsec$ pixels for display; the contours are
separated by a factor of 2 in intensity and the direction of Celestial
North is up and East to the left.  The image has been corrected for
the effects of exposure variations and telescopic vignetting. The
image has been smoothed for visual clarity with a Gaussian of $\sigma
= 1$ pixel, although the image used for analysis is not smoothed as
such.  

\plottwo{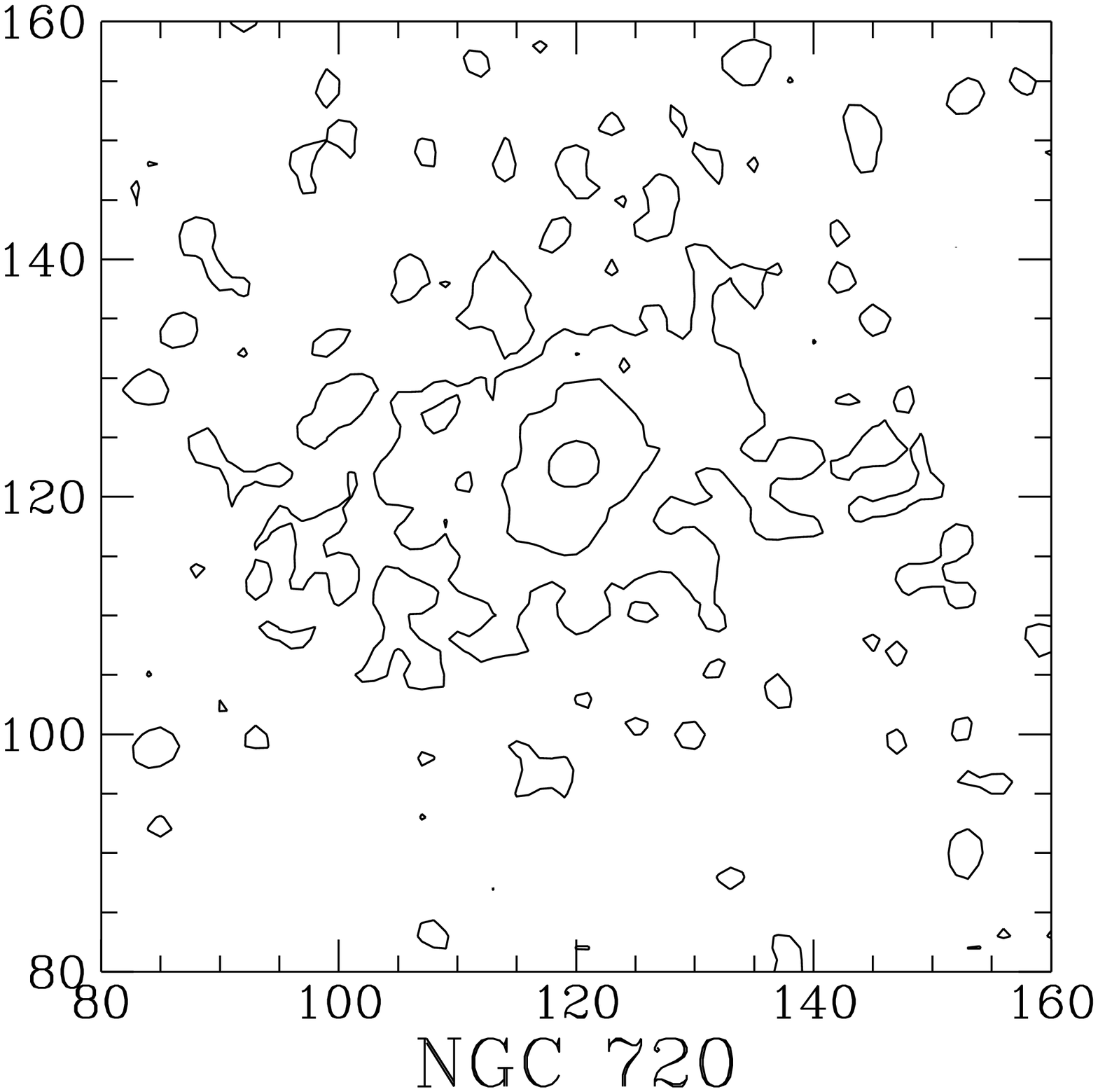}{blank.ps}

\end{figure}

The reduced image is plotted in Figure \ref{fig.image}, where for
display we have rebinned the image into $5\arcsec$ pixels and smoothed
with a Gaussian of $\sigma=1$ pixel.

The final step in the image reduction is to subtract the
background. We only require background subtraction for computation of
the radial profile, not for computation of the ellipticities and
position angles.  Since we extend the radial profile into regions
where the galaxy flux $\lesssim$ background level we need to carefully
estimate the background to reduce large fractional systematic errors
in the radial profile at large radii.  The HRI surface brightness
profile of NGC 720 reaches the background level for $r\sim
300\arcsec$, somewhat smaller than the outer radius $r\sim 400\arcsec$
we detected with the more sensitive PSPC (BC).

To estimate the background level we compared circular annuli of
different radii centered on NGC 720. We restricted analysis to regions
outside of $r=400\arcsec$ to minimize residual contribution from the
galaxy. Moreover, to reduce any errors in the HRI vignetting
correction and possible scattered light in the outer regions of the
HRI image we confined our analysis to regions as near the center of
the field as possible.  From consideration of these issues we adopted
the $500\arcsec - 600\arcsec$ annulus to estimate the background level
where the residual contribution to the emission from NGC 720 should be
$<1\%$. We estimate that this background level is accurate to within
$\sim 3\%$.

\section{X-ray Isophote Shapes and Orientations \label{iso}}

As is typical for X-ray images of early-type galaxies the small number
of counts $(\sim 1000)$ for the HRI image of NGC 720 implies that we
can only hope to measure with any precision the ellipticity and
position angle of the aggregate X-ray surface brightness in a large
aperture. The method we employ is an iterative procedure and is
analogous to computing the two-dimensional moments of inertia within
an elliptical region where the ellipticity, $\epsilon_M$, is given by
the square root of the ratio of principal moments and the orientation
of the principal moments gives the position angle, $\theta_M$ (see BC;
Buote \& Canizares 1996). The parameters $\epsilon_M$ and $\theta_M$
are good estimates of the ellipticity $(\epsilon)$ and position angle
($\theta$, or PA) of an intrinsic elliptical distribution of constant
shape and orientation. For a more complex distribution $\epsilon_M$
and $\theta_M$ are average values weighted heavily by the outer parts
of the regions.

To estimate the uncertainties on $\epsilon_M$ and $\theta_M$ we employ
a Monte Carlo procedure described by Buote \& Tsai (1996) slightly
modified for application to HRI images.  Because the HRI surface
brightness profile of NGC 720 appears to be rather complex in shape
and orientation, we apply the Monte Carlo procedure directly to the
HRI image rather than some best-guess model (e.g., an elliptical
$\beta$ model) which may not account for all of the relevant features.
To construct an ``average'' model for the real HRI image we smooth the
reduced image (including background) with a Gaussian of width $\sigma
= 2.5\arcsec$ (approximately half the width of the on-axis HRI PSF) to
better approximate pixels with $0$ counts in the real image.  We
examined the sensitivity of the derived values of $\epsilon_M$ and
$\theta_M$ to the value of $\sigma$.  The estimated $68\%$ confidence
limits on $\epsilon_M$ and $\theta_M$ for semi-major axes $\gtrsim
30\arcsec$ typically vary by $<10\%$ for values of $\sigma = 0\arcsec
- 2.5\arcsec$.  For $\sigma=5\arcsec$ the derived confidence estimates
for these semi-major axes vary by as much as 25\%. However, for
semi-major axes $\gtrsim 75\arcsec$ the $68\%$ confidence limits vary
by $<5\%$ for $\sigma=0\arcsec - 5\arcsec$.

To the smoothed image we added point sources having spatial properties
consistent with the HRI PSF and numbers consistent with the $\log
N(>S)$ - $\log S$ distribution given by Hasinger et al. (1993).  We
excluded from the simulations any bright point sources
(i.e. $>5\sigma$ above the background) appearing within $\sim
100\arcsec$ of the galaxy center as they would be easily detected by
visual examination and removed. Poisson noise was then added to the
image consistent with the exposure of the NGC 720 HRI observation.

We performed 1000 realizations and defined the 90\% (for example)
confidence limits on $\epsilon_M$ and $\theta_M$ for each semi-major
axis to be the 50th smallest and the 50th largest values obtained from
the 1000 simulations. Although this definition is arbitrary it provides
a simple, realistic measure of the significance of the ellipticity and
PA. 

In Table \ref{table.e0pa} we list the results for $\epsilon_M$ and
$\theta_M$ and their associated 68\% and 90\% confidence estimates for
semi-major axes $(a)$ ranging from $10\arcsec$ to $120\arcsec$. The
isophotes are significantly elongated with $\epsilon_M\sim 0.3 \pm
0.1$ for $10\arcsec \le a\le 120\arcsec$. These values for
$\epsilon_M$ are consistent with the PSPC values (BC) for $a\gtrsim
75\arcsec$ but are systematically larger than those obtained with the
PSPC for $a\lesssim 75\arcsec$, which is consistent with the PSPC
values at small $a$ being reduced as a result of its larger PSF.

\begin{figure}[p]
\caption{  \label{fig.pa} }
\raggedright

The HRI position angle computed within a specified semi-major axis,
$a$, and the estimated 95\% confidence intervals; note the position
angles are correlated since each $a$ contains the regions $<a$. The
optical and PSPC $PA$ are shown approximately over their relevant
range of validity.

\plottwo{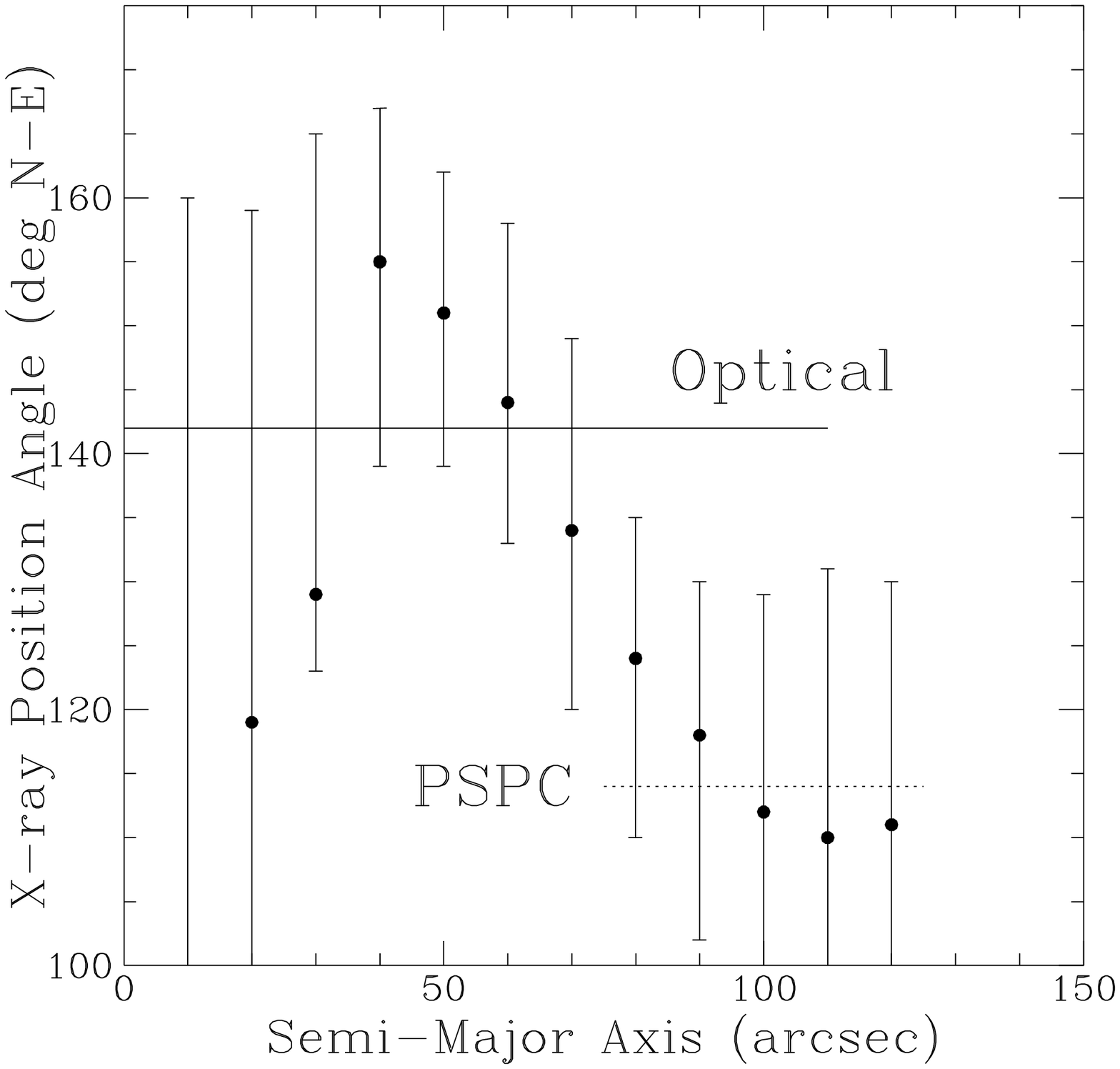}{blank.ps}

\end{figure}

The position angles of the isophotes vary significantly with
semi-major axis; the $PA$ profile and the estimated $2\sigma$ errors
are plotted in Figure \ref{fig.pa}. For $a\lesssim 60\arcsec$, the
$PA$ is consistent with the optical $PA$ of $142\arcdeg$ (e.g.,
Peletier et al. 1990). At larger $a$, however, the PA decreases and
levels off at values consistent with $PA=114\arcdeg$, the position
angle obtained from analysis of the PSPC at these $a$. The variation
is highly significant: for $a=90\arcsec$ we estimate that
$PA=99\arcdeg - 132\arcdeg$ at the 95\% confidence level, and
$PA=90\arcdeg - 137\arcdeg$ at the 99\% level, still significantly
less than the measured value of $144\arcdeg$ at $a=60\arcsec$.

Possible origins for this $PA$ twist are chance alignments of a few
appropriately placed unresolved foreground and background sources, the
residual effects of ram pressure distortions, or tidal interactions
with neighboring galaxies.  For the case of tidal fields, there are no
giant galaxies within 1 Mpc [Dressler, Schechter, \& Rose 1986] which
means that the expected gravitational field due to a neighbor
comparable in size to NGC 720 would contribute only 1\% to the
potential at a radius of 10 kpc from NGC 720.  If the other effects
are important then the surface brightness should display substantial
departures from elliptical symmetry.  To investigate this possibility
we first examined whether the centroid of the surface brightness
shifts with increasing aperture size. Over the whole range of $a$
explored the centroid shifts by less than one pixel in both $x$ and
$y$ directions; i.e. there is no significant centroid shift.

We also examined higher order symmetry. First, we aligned coordinate
axes centered on the galaxy with the $x$ axis at the optical
$PA=142\arcdeg$. Within a radius of $r=60\arcsec$ we computed the
counts in the four quadrants in these coordinates. The results are
listed in Table \ref{table.sym}. The counts are consistent with a mean
value of $\sim 285$ and their $1\sigma$ Poisson uncertainties of $22$;
i.e. within $r=60\arcsec$ there is no evidence for asymmetry between
the four quadrants. By instead aligning the coordinate axes with
$PA=114\arcdeg$ and examining the surface brightness in an annulus
from $r=60\arcsec - 90\arcsec$ we explored the symmetry in the outer
regions; see Table \ref{table.sym}. Again, there is no evidence for
asymmetries.

Unresolved point sources (and statistical noise) cannot produce the
observed $PA$ twist as demonstrated by the above Monte Carlo
simulations (Table \ref{table.e0pa}).  To further investigate this
issue we estimated the probability that unresolved sources contribute
sufficiently to (specifically) the outer, low surface brightness,
isophotes to produce the position angle twist.  The region of interest
lies between the isophotes with $a\sim 75\arcsec - 100\arcsec$.  If
part of this region is significantly contaminated by point-source
emission, then a $PA$ twist could result. Consider, e.g., a
$25\arcsec\times 25\arcsec$ square within this region in which the
flux from NGC 720 is measured to be $\sim 4\times 10^{-15}$ erg
cm$^{-2}$ s$^{-1}$ in the 0.1-2.4 keV band using the PSPC image.  From
the $\log N(>S) - \log S$ distribution for ROSAT X-ray sources
(Hasinger et al. 1993) we estimate the probability that there is a
source with flux at least half of the measured value in the square
region to be only $\sim 2\%$. Hence, it is unlikely that the gross
features of the surface brightness in the key regions $r\sim 75\arcsec
- 100\arcsec$ are affected substantially by the emission from
unresolved foreground and background sources.

\section{Radial Profile \label{radpro}}

Using the background-subtracted HRI image we constructed the
azimuthally averaged radial profile located at the centroid of the
galaxy emission. The centroid was computed in a circular aperture of
radius $120\arcsec$ which effectively encloses $\sim 85\%$ of the
total flux of the galaxy; note that the centroid varies by
$<2.5\arcsec$ for smaller aperture sizes. The radial profile was
rebinned so that each bin had $S/N \ge 6$. This yields six $5\arcsec$
bins from $r=0\arcsec - 30\arcsec$, three $15\arcsec$ bins from
$r=30\arcsec - 75\arcsec$, and one $45\arcsec$ bin from $r=75\arcsec -
120\arcsec$. The radial profile is plotted in Figure \ref{fig.radpro}.

For $r\gtrsim 30\arcsec$, the radial profile is consistent with a
power-law form.  However, the profile appears to level off between
$r\sim 10\arcsec - 30\arcsec$ and then rises again for $r\lesssim
10\arcsec$. These qualitative features of the radial profile appear to
be robust to small changes in the sizes of the radial bins.  As we
show in the next section, although these features are not of
sufficient $S/N$ to distinguish between a wide range of models, they
are of sufficient quality to eliminate some of the simplist models
that previously were consistent with the X-ray data for NGC 720.

\section{Analysis and Discussion \label{anal}}

Since the emission from unresolved foreground and background sources
and effects from ram pressure and tidal distortions appear to be
insignificant (see \S \ref{iso}), we now consider origins of the
surface brightness features that are intrinsic to the galaxy.  First
we examine whether the HRI data may be described by a model where the
X-ray emission is entirely due to hot gas in hydrostatic equilibrium
with the underlying potential of NGC 720. We shall consider two
classes of models -- $\beta$ models and power-law or Hernquist
Spheroidal Mass Distributions (SMDs). First we take the surface
brightness distribution to be given by the $\beta$ model (Cavaliere \&
Fusco-Femiano 1976), $\Sigma_{hg}(R) = \Sigma_0[1 +
(R/R_c)^2]^{-3\beta+1/2}$, which is a power law for $R\gg R_c$.  This
model corresponds to a logarithmic potential at large radii (e.g.,
Appendix C. of Trinchieri et al. 1986) and is generally a good fit to
the X-ray radial profiles of ellipticals observed with the {\it
Einstein} IPC (e.g., Forman, Jones, \& Tucker 1985; Trinchieri et
al. 1986) and the ROSAT PSPC -- in particular that of NGC 720 (see
BC).

To perform a more consistent comparison between the model and data we
evaluate the model on a grid of pixels corresponding to the same scale
as the data and convolve the model with the relevant PSF. The radial
profile is then constructed and binned in the same manner as done for
the data. This procedure does yield slightly different parameter
values and $\chi^2$ values than if the convolved model is directly
compared to the radial profile of the data as is typically done. In
all of the fits we set the ellipticity of the hot gas to $\epsilon_x =
0.25$ which is essentially the mean value of the X-ray isophotes. The
fitted parameters and $\chi^2$ values change by $\lesssim 5\%$ when
considering ellipticities $\epsilon_x = 0.2-0.3$.

The $\beta$ model is a poor fit to the HRI radial profile: $\chi^2=18$
for 7 dof. The poor quality of the fit is due entirely to the inner
radial bins $(r\le 30\arcsec)$. In contrast, BC showed that the
$\beta$ model provides a good fit to the PSPC profile of NGC 720.  We
reanalyzed the PSPC data and rebinned the PSPC radial profile so that
each bin has $S/N>5$ and energy range 0.5-2.0 keV; BC instead fixed
the bin size to $15\arcsec$.  The $\beta$ model applied to this
rebinned PSPC radial profile yields a formally marginal fit:
$\chi^2=19$ for 10 dof. Similar to the HRI data, the major source of
discrepancy between model and data is due to the two inner bins
($0\arcsec - 15\arcsec, 15\arcsec - 30\arcsec)$ -- the bin near
$90\arcsec$ is not as important due to its larger fractional
statistical uncertainty. The best-fit parameters for the HRI and PSPC
fits are listed in Table \ref{fit}.  The HRI profile is shallower
(i.e. smaller $\beta$) than that of the PSPC which is probably an
artifact of the weak constraints at large radii ($r\gtrsim
100\arcsec$); i.e. when the PSPC data is fit only to bins within
$r=100\arcsec$ then smaller $\beta$ values are also obtained.

\begin{figure}[p]
\caption{  \label{fig.radpro} }
\raggedright

(a) The HRI background-subtracted radial profile binned such that each
bin has $S/N\ge 6$. The solid line represents the best-fit $\beta$
model and the dotted line corresponds to the best-fit $\beta$ +
discrete model.  The midpoint of each bin is used to define its
location in the plot; note the lines are only defined at the midpoints
of the bins. The dashed horizontal lines represent the background
levels. (b) The PSPC background-subtracted radial profile binned such
that each bin has $S/N > 5$. The squares and crosses represent the
corresponding best-fit models from (a).

\plottwo{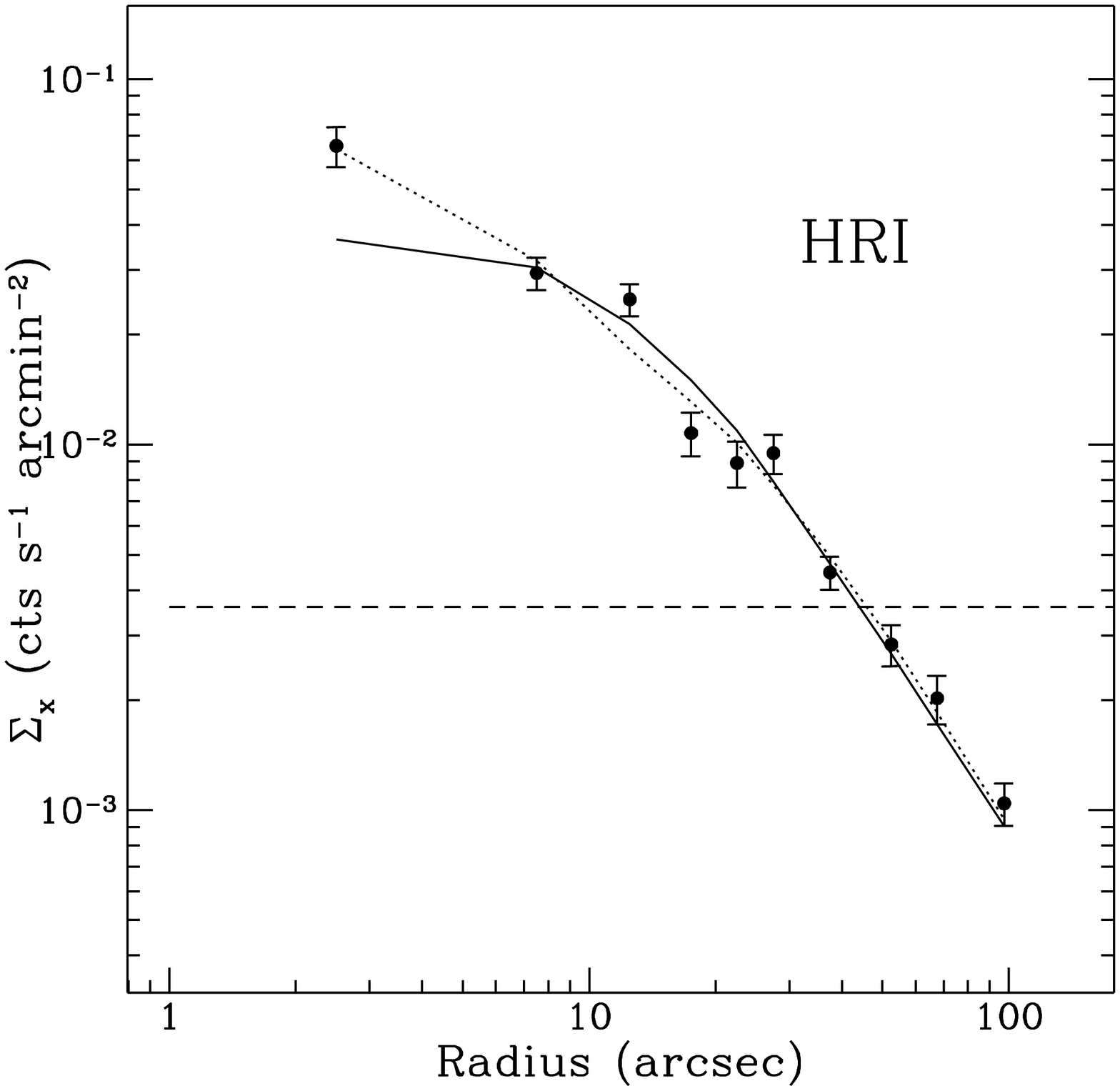}{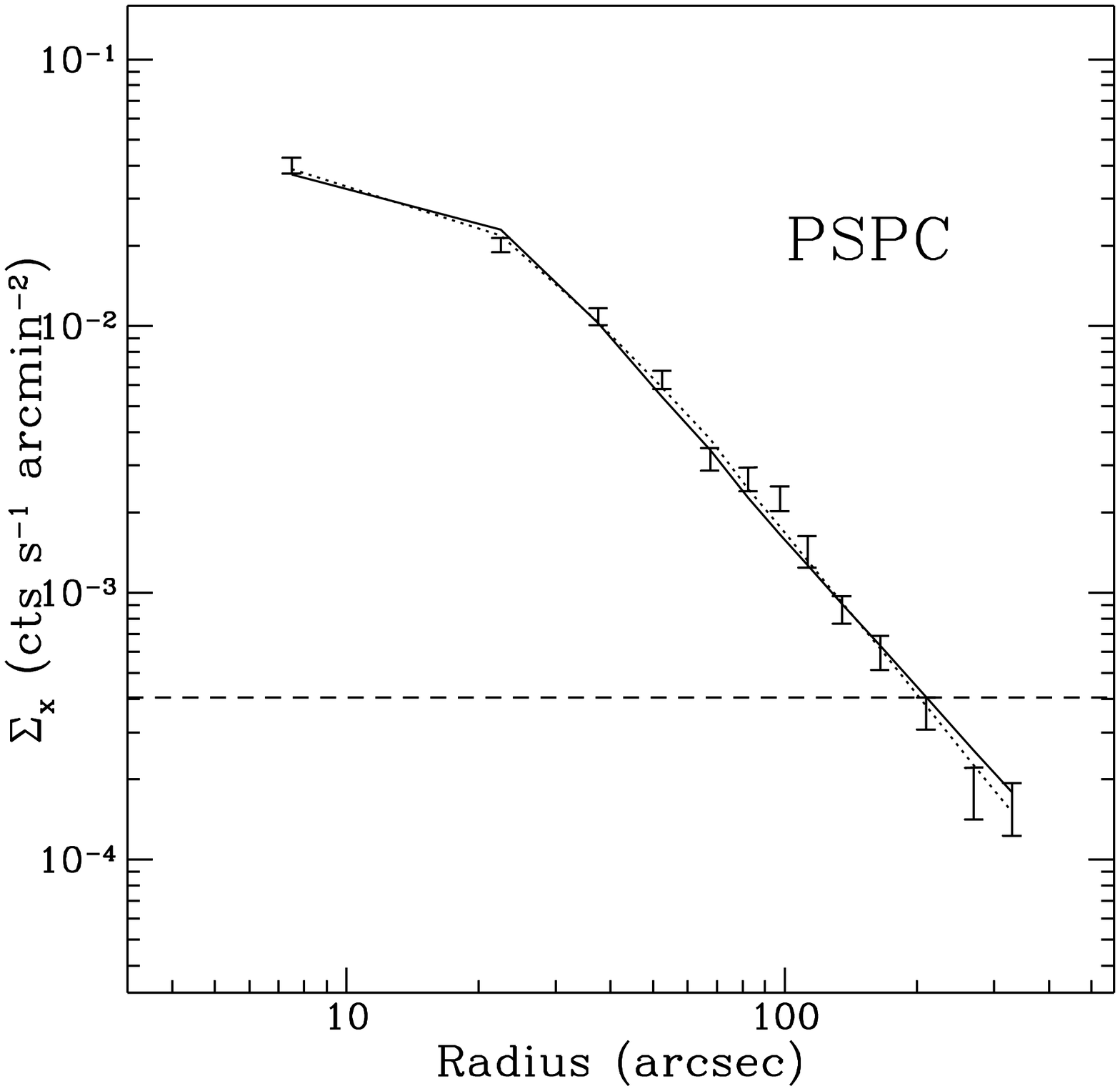}

\end{figure}

It is necessary to fit the model jointly to the HRI and PSPC data in
order to obtain constraints acceptable for both data sets; the model
for each data set is convolved with its appropriate PSF. By doing this
we now have four free parameters: $a_x$, $\beta$, and the
normalizations for the HRI and PSPC radial profiles. The joint fit is
shown in Figure \ref{fig.radpro} and the best-fit and 90\% confidence
levels (2 interesting parameters) are shown in Table \ref{fit}. The
fit is formally worse ($\chi^2=53$ for 19 dof) than the fits to the
data sets individually which is primarily due to joining the shallower
HRI profile to the steeper PSPC profile; note the poor fit is not the
result of the small ($<$ few \%) uncertainties in the background
estimates.

These qualitative and quantitative results are reproduced when we
consider the spheroidal mass models (SMDs) used in BC that are
consistent with the PSPC spatial and spectral (i.e. nearly isothermal)
data; i.e. models where the mass density is stratified on concentric,
similar spheroids and has either a power-law or Hernquist (1990)
radial profile. For example, a model with mass density $\rho\sim
r^{-2}$ and an ellipticity of 0.50 fit jointly to the HRI and PSPC
data yields a poor fit: $\chi^2=46$ for 19 dof. Similar results are
obtained for Hernquist (1990) models and its generalization to
shallower profiles (i.e. $\rho\sim r^{-1}(a_c + r)^{-1}$).  Hence,
simple plausible models for the hot gas cannot fit the radial profiles
of the HRI and PSPC data, where the disagreement between models and
data primarily occurs in the inner regions $r\lesssim 30\arcsec$.

We are thus compelled to increase the complexity of our model to fit
the X-ray surface brightness. There must be significant contribution
to the X-ray emission from discrete sources in NGC 720; i.e., the
integrated emission from X-ray binaries (see, e.g., Canizares,
Fabbiano, \& Trinchieri 1987; Kim, Fabbiano, \& Trinchieri 1992). This
``discrete'' component should have a spectrum that is harder than that
of the hot gas and essentially similar to that found in spirals where
discrete sources are likely to account for most of the emission.

Considering the relatively large value of $L_X/L_B$ for NGC 720 we do
not expect a large contribution to the X-ray emission from discrete
sources in the 0.1 - 2.4 keV energy range (Kim et al. 1992). We
estimate the relative flux of the hot gas and a harder component using
the PSPC spectrum computed in a circle of $120\arcsec$ about the
galaxy centroid. We fitted the spectrum with two Raymond-Smith (1977,
updated to current version) models where we fixed the column density
to its Galactic value (Stark et al. 1992), the abundances to 20\%
solar (approximately the best-fit single-component value -- see BC),
and the temperature of the hard component to 4.5 keV which corresponds
to the integrated spectrum of Sb+Sc galaxies analyzed by Kim et
al. (1992). This temperature is also consistent with the hard
component determined for NGC 4472 with ASCA (Awaki et al. 1994;
Matsushita et al. 1994), a galaxy that has a value of $L_X/L_B$ very
similar to NGC 720. We find for these models that the emission of the
hard component can make up at most 32\% of the total emission within
$120\arcsec$ (90\% confidence on two interesting parameters) but with
a lower limit approaching zero (0.1\%). This range is consistent with
the ASCA observation of NGC 4472 where Awaki et al. (1994) determined
that only $\sim 10\%$ of the X-ray emission in the 0.5 - 5 keV band is
due to a hard component (for about twice the radius used for NGC
720). In the 0.1-2.4 keV band of the HRI and PSPC this percentage
should be even smaller.

We place alternative constraints on the total emission from discrete
sources using the spatial properties of the X-ray surface
brightness. Although there exists no strong observational constraints
on the spatial distribution of such a component in ellipticals, we
make the reasonable assumption that the discrete emission has a
spatial distribution similar to that of the optical light.  We take
the surface brightness of the discrete component, $\Sigma_{ds}$, to
have the two-dimensional distribution of the $B$-band light of NGC
720: the HST data from Lauer et al. (1995) define the core while
the rest of the galaxy is defined by the data from Peletier et
al. (1990), who find no evidence for color gradients in NGC 720.  The
hot-gas component, $\Sigma_{hg}$, is again modeled as either the
$\beta$ or the isothermal power-law or Hernquist (1990) SMDs.

In order to specify the parameters of this discrete + hot gas model we
fitted this model jointly to radial profiles of the HRI and PSPC data,
convolved with the appropriate PSFs, and using the same procedure as
above. The HRI data provide the dominant constraint at small radii
$(R\lesssim 30\arcsec)$ while the PSPC data dominates over the rest of
the galaxy $(R\sim 30\arcsec - 400\arcsec)$.  For each fit the ratio
of emission from hot gas to the emission from discrete sources,
$\Sigma_{hg}/\Sigma_{ds}$, is computed in a circle of radius
$120\arcsec$ from the galaxy centroid.

The results of the joint fits to the HRI and PSPC radial profiles are
listed in Table \ref{fit} and shown in Figure \ref{fig.radpro}.  Let
us first consider the case where the hot gas is taken to be a $\beta$
model. The range of ratios of discrete emission to hot gas obtained
are consistent with the above spectral results from the PSPC which
serves as a comforting consistency check between the spatial and
spectral methods.  Although only one free parameter has been added
$(\Sigma_{hg}/\Sigma_{ds})$ to our model, the quality of the fit is
substantially improved ($\chi^2 = 29$ for 18 dof), though still
formally only a marginal fit ($P_{\chi^2}=95\%$).  Again the
discrepancy between model and data may be attributed primarily to the
inner $30\arcsec$ region.  The model fits the PSPC profile acceptably,
but the HRI profile (particularly the core) is poorly fit:
$\chi^2_{PSPC}=13$ for 9 dof and $\chi^2_{HRI}=15$ for 6 dof. In
contrast to the hot gas alone, adding the discrete component allows
the innermost bin to be well fitted (i.e. this combined hot gas +
discrete model can account well for the excess emission in the central
bin). However, this model does not produce the flattening of the HRI
radial profile between $r\sim 10\arcsec - 30\arcsec$, which is the
dominant source of discrepancy between model and data. When the hot
gas is modeled using the SMDs (see above) we obtain results in
excellent agreement with the $\beta$ models (see Table \ref{fit}).

Although the formal fit to the data is marginal, it is worth examining
whether the discrete + hot gas model can reproduce the observed
position angle twist. We take the discrete model to be oriented along
the optical PA while the hot gas is oriented along the PA of the outer
isophotes $(PA = 114\arcdeg)$. We find that the models do not produce
large PA twists, the largest twist is generally produced by the
best-fit model to the radial profile where $\Delta PA\sim 10\arcdeg$
between semi-major axes of $60\arcsec$ and $100\arcsec$.  The
statistical uncertainties on simulated observations of these models
are rather large and similar to what was found for the data (Table
\ref{table.e0pa}): the observed PA twist between semi-major axes of
$60\arcsec$ and $100\arcsec$ is inconsistent at the estimated $\sim
90\%$ confidence level.  Since, however, an acceptable model must
produce the $PA$ {\it profile} between $60\arcsec$ and $100\arcsec$
(not simply the endpoints), the discrepancy may be more
pronounced. (This issue requires significantly more computational
effort to properly address than is appropriate for our present study.)
Hence, the discrete + hot gas model cannot fully produce the observed
large PA twist (though it will contribute) or the features in the core
of the radial profile.

Joint consideration of the HRI and PSPC data of NGC 720 suggests that
the X-ray emission is more complex than simple isothermal spheroidal
models of the hot gas combined with emission from discrete sources.
Detailed investigation of more sophisticated models is beyond the
scope of this paper, but we briefly describe some possible solutions.

{\it Triaxiality:} A PA twist can be produced by the projection of a
triaxial distribution whose three-dimensional isodensity surfaces have
shapes which vary as a function of radius (e.g., Mihalas \& Binney
1981).  The ellipticities of the optical isophotes do not vary
substantially over the region of interest ($a\gtrsim 30\arcsec$; e.g.,
Peletier et al. 1990). However, the stars do not trace the same shape
as the potential (because of anisotropic velocity dispersions) and
thus do not in general have the same three-dimensional spatial
distribution, and thus projected distribution, as the gas.  N-body
simulations of galaxies and clusters generically produce strongly
triaxial halos (e.g., Frenk et al. 1988; Dubinski \& Carlberg
1991). In fact, hydrodynamic simulations often produce halos that are
nearly oblate in the central regions while the outer parts,
particularly for clusters formed in filaments, are nearly prolate
(e.g., Katz \& Gunn 1992; Dubinski 1994; Buote \& Tsai 1995). Perhaps
a large observed PA twist can result if the transition from oblate
($r\lesssim 60\arcsec$) to prolate ($r\gtrsim 100\arcsec$) is
sufficiently rapid. However, using simple models with the above
specifications and a power-law variation in axial ratio between $a\sim
60\arcsec$ and $a\sim 100\arcsec$ we find it difficult to produce $PA$
twists larger than $\sim 15-20\arcdeg$ which also produce the observed
ellipticities of the isophotes.

{\it Intrinsic Misalignment of Light and DM Halo:} If the principal
axes of the stars and dark matter halo are intrinsically misaligned
then a $PA$ twist would naturally occur. Such a scenario has been
proposed by Ostriker \& Binney (1989) to explain the origin of warps
in disk galaxies. However, Dubinski \& Kuijken (1995) have shown that
large misalignments will rapidly decay because the disk, with its
large spin misaligned with the halo, will precess and suffer dynamical
friction with the halo, and thus settle into the halo equatorial
plane; often the inner halo actually aligns with the disk because the
disk is difficult to re-orient due to its large spin. Because a giant
elliptical like NGC 720 does not rotate substantially
($v/\sigma^{\ast} = 0.15$, e.g., Fried \& Illingworth 1994), it is not
clear whether these arguments imply the same short timescales. Perhaps
another mechanism (e.g., phase mixing) ensures alignment for an
elliptical. Even if the timescale for alignment is found to be short
compared to a Hubble time (thus ruling out a primordial misalignment),
it may be that the misalignment is caused by a dwarf companion that
has recently struck NGC 720; there are several known dwarf companion
galaxies around NGC 720 (Dressler et. al. 1986).

{\it Rotating Cooling Flow:} The flattening of the radial profile
between $r\sim 10\arcsec - 30\arcsec$, though not of high enough
significance to warrant detailed modeling, deserves some attention. It
is possible that the surface brightness features on these small scales
are due to patchy absorption by cold clouds condensed out of the hot
phase (e.g., Fabian 1994). However, another aspect of the cooling gas
may be the origin for the features.  Even though giant ellipticals
like NGC 720 typically rotate slowly, in principle slowly rotating gas
at large galacto-centric radii can cool, flow towards the center,
eventually be driven to dynamically important rotational velocities if
angular momentum is conserved (e.g., Cowie, Fabian, \& Nulsen 1980),
and thereby form a spinning disk (e.g., Kley \& Mathews 1995).  Not
only will the ellipticity of the X-ray isophotes become more elongated
in the region of dynamically important rotation, the shape of the
radial profile will be affected as well (e.g., Kley \& Mathews 1995).

\section{Conclusions \label{conc}}

The interesting features of the HRI surface brightness of NGC 720
provide more questions than answers into the origin of its X-ray
emission and the distribution of its underlying mass. Presently ASCA,
with its improved spectral resolution over ROSAT, can help elucidate
the role of the total emission from discrete sources in NGC
720. However, we must await the launch of AXAF to have the combined
spatial and spectral resolution to distinguish models of the origin of
the position-angle twist and flattening of the radial profile of the
HRI data.

\acknowledgements

We thank J. Dubinski, A. Fabian, L. Hernquist, W. Mathews, P. Sackett,
P. Schechter, D. Syer, J. Tonry, S. Tremaine, and S. Zepf for
insightful discussions.  We specially thank D. Harris and others at
hotseat@cfa.harvard.edu for advice regarding the reduction of the HRI
data.  This research was supported by grants NASGW-2681 (through
subcontract SVSV2-62002 from the Smithsonian Astrophysical
Observatory), and NAG5-2921.

\vfill\eject

\begin{table}[p]
\caption{X-ray Ellipticities and Position Angles\label{table.e0pa}}
\begin{tabular}{rrrrrrrrr} \tableline\tableline
\multicolumn{1}{c}{$a$} \\
\multicolumn{1}{c}{(arcsec)} && \multicolumn{1}{c}{$\epsilon_M$} &
\multicolumn{1}{c}{68\%} & \multicolumn{1}{c}{90\%} &
\multicolumn{1}{c}{$\theta_M$} & \multicolumn{1}{c}{68\%} &
\multicolumn{1}{c}{90\%} & \multicolumn{1}{c}{Counts}\\ \tableline
10$\ldots\ldots$  && 0.36 & 0.06 - 0.37 & 0.02 - 0.51 & 75 & 45 -
132 & 19 - 160 & 158\\
20$\ldots\ldots$  && 0.14 & 0.08 - 0.28 & 0.03 - 0.36 & 119 & 84 -
144 & 60 - 159 & 418\\
30$\ldots\ldots$  && 0.39 & 0.22 - 0.39 & 0.16 - 0.44 & 129 & 131
- 156 & 123 - 165 & 527\\
40$\ldots\ldots$  && 0.35 & 0.25 - 0.40 & 0.20 - 0.45 & 155 & 145
- 162 & 139 - 167 & 720\\
50$\ldots\ldots$  && 0.28 & 0.25 - 0.37 & 0.20 - 0.41 & 151 & 144
- 157 & 139 - 162 & 908\\
60$\ldots\ldots$  && 0.35 & 0.23 - 0.38 & 0.17 - 0.40 & 144 & 139
- 153 & 133 - 158 & 977\\
70$\ldots\ldots$  && 0.30 & 0.24 - 0.38 & 0.18 - 0.44 & 134 & 126
- 144 & 120 - 149 & 1130\\
80$\ldots\ldots$  && 0.33 & 0.25 - 0.39 & 0.20 - 0.42 & 124 & 114
- 130 & 110 - 135 & 1220\\
90$\ldots\ldots$  && 0.28 & 0.23 - 0.36 & 0.18 - 0.41 & 118 & 108
- 125 & 102 - 130 & 1351\\
100$\ldots\ldots$ && 0.28 & 0.17 - 0.34 & 0.12 - 0.38 & 112 & 103
- 123 & 95 - 129 & 1427\\
110$\ldots\ldots$ && 0.24 & 0.19 - 0.35 & 0.12 - 0.39 & 110 & 99 -
124 & 87 - 131 & 1537\\
120$\ldots\ldots$ && 0.28 & 0.18 - 0.35 & 0.09 - 0.40 & 111 & 97 -
121 & 89 - 130 & 1611\\ \tableline  

\end{tabular}

\tablecomments{The values of $\epsilon_M$ (and confidence limits) are
computed within an aperture of semi-major axis $a$ on the image having
$2.5\arcsec$ pixels with the background included; the counts, however,
have the background subtracted.}

\end{table}

\begin{table}[p]
\caption{Symmetry of X-ray Surface Brightness\label{table.sym}}
\begin{tabular}{|c|c|c|} \tableline\tableline
& $PA = 142\arcdeg$ & $PA = 114\arcdeg$ \\
Quadrant & $(r=60\arcsec)$ & $(60\arcsec\le r\le 90\arcsec)$ \\
\tableline 
I   & $303 \pm 23$  & $86 \pm 18$\\           
II  & $253 \pm 21$  & $96 \pm 18$\\           
III & $287 \pm 22$  & $82 \pm 18$\\           
IV  & $284 \pm 22$  & $99 \pm 18$\\ \tableline
\end{tabular}

\tablecomments{Background-subtracted counts and their associated
$1\sigma$ uncertainties assuming Poisson statistics. See \S \ref{iso}
for an explanation of the regions.}

\end{table}

\begin{table}[p]
\caption{Radial Profile Models\label{fit}}
\begin{tabular}{lccccc} \tableline\tableline
& $R_c$ \\
Model & (arcsec) & $\beta$ & $\Sigma_{ds}/\Sigma_{hg}$ & $\chi^2$ &
dof\\ \tableline
$\beta$ - HRI & $6\pm 3$ & $0.40\pm 0.03$ & $\ldots$ & 18
& 7\\
$\beta$ - PSPC & $18\pm 4$ & $0.49\pm 0.02$ & $\ldots$ & 19
& 10\\ 
$\beta$ - BOTH & $13\pm 3$ & $0.46\pm 0.02$ & $\ldots$ & 53
& 19\\
$\beta$ + DISCRETE & $30^{+16}_{-8}$ & $0.50^{+0.05}_{-0.01}$ &
$0.40^{+0.27}_{-0.20}$ & 29 & 18\\ 
SMD - BOTH &$\ldots$ & $\ldots$ & $\ldots$ & 46 & 19 \\
SMD + DISCRETE & $\ldots$ &$\ldots$ & $0.40^{+0.20}_{-0.27}$ & 29 &
18\\ \tableline 
\end{tabular}

\tablecomments{Under ``Models'', $\beta$ stands for the hot gas
represented by the $\beta$ model while SMD indicates the spheroidal
mass models found by BC to be consistent with the PSPC data for NGC
720. ``BOTH'' indicates model is fit jointly to HRI and PSPC data. The
DISCRETE models are fit jointly to both data sets. For the SMD models
we only list the value of $\chi^2$ and the flux ratio for the DISCRETE
case.}

\end{table}


\clearpage

\clearpage
\begin{references}

\reference{A} Aschenbach, B. 1988, \ao, 27, No. 8, 1404

\reference{A} Awaki, H. et al. 1994, \pasj, 46, L65

\reference{A} Birkinshaw, M., \& Davies, R. L. 1985, \aj, 291, 32

\reference{A} Buote, D. A., \& Canizares, C. R. 1994, \apj, 427, 86
(BC) 

\reference{A} Buote, D. A., \& Canizares, C. R. 1996, \apj, 457, 177

\reference{A} Buote, D. A., \& Tsai, J. C. 1995, \apj, 439, 29

\reference{A} Buote, D. A., \& Tsai, J. C. 1996, \apj, 458, 27

\reference{A} Canizares, C. R., Fabbiano, G., \& Trinchieri, G. 1987,
\apj, 312, 503

\reference{A} Cavaliere, A., \& Fusco-Femiano, R. 1976, \aap, 49, 137

\reference{A} Cowie, L. L., Fabian, A. C., \& Nulsen, P. E. J. 1980,
191, 399 

\reference{A} David, L. P., Harnden, F. R., Kearns, K. E., \& Zombeck,
M .V. 1995, The ROSAT High Resolution Imager (HRI), (U.S. ROSAT Science
Data Center -- Smithsonian Astrophysical Observatory)

\reference{A} de Zeeuw, T., in 36th Herstmonceaux Conf., Gravitational
Dynamics, ed. O. Lahav, E. Terlevich, \& R. Terlevich (Cambridge:
Cambridge Univ. Press), in press (astro-ph/9510145)

\reference{A} de Zeeuw, T., \& Franx, M. 1991, \araa, 29, 239

\reference{A} Dressler, A., Schechter, P. L., \& Rose, J. A. 1986, \aj,
91, 1058 

\reference{A} Dubinski, J. 1994, \apj, 431, 617

\reference{A} Dubinski, J., \& Carlberg, R. 1991, \apj, 378, 496

\reference{A} Dubinski, J., \& Kuijken, K. 1995, \apj, 442, 492

\reference{A} Fabian, A. C. 1994, \araa, 32, 277

\reference{A} Forman, W., Jones, C., \& Tucker, W. 1985, \apj, 293,
102 

\reference{A} Frenk, C. S., White, S. D. M., Davis, M., Efstathiou,
G. 1988, \apj, 327, 507

\reference{A} Fried, J. W., \& Illingworth, G. D. 1994, \aj, 107, 992

\reference{A} Hasinger, G., Burg, R., Giacconi, R., Hartner, G.,
Schmidt, M., Tr\"{u}mper, J., \& Zamorani, G. 1993, \aap, 275, 1

\reference{A} Hernquist, L. 1990, \apj, 356, 359

\reference{A} Katz, N., \& Gunn, J. E. 1991, \apj, 377, 365

\reference{A} Katz, N., \& White, S. D. M. 1993, \apj, 412, 455 

\reference{A} Kim, D.-W., Fabbiano, G, \& Trinchieri, G. 1992, \apjs, 80,
645

\reference{A} Kley, W., \& Mathews, W. G. 1995, \apj, 438, 100

\reference{A} Lauer, T. R., Ajhar, E. A., Byun, Y., Dressler, A., \&
Faber, S. M. 1995, \aj, 110, 2622

\reference{A} Matsushita, K., et. al. 1994, \apj, 436, L41

\reference{A} Mihalas, D., \& Binney, J. 1981, Galactic Astronomy (New
York: Freeman)

\reference{A} Morse, J. A. 1993, in The Soft X-ray Cosmos (AIP
Conf. Series 313), ed. E. M. Schlegel, \& R. Petre (AIP Press: New
York), 252

\reference{A} Ostriker, E. C., \& Binney, J. J. 1989, \mnras, 237, 785

\reference{A} Peletier, R. F., Davies, R. L., Illingworth, G. D.,
Davis, L. E., Cawson, M. 1990, \aj, 100, 1091

\reference{A} Raymond, J. C., \& Smith, B. W. 1977, \apjs, 35, 419

\reference{A} Sackett, P. D., Gravitational Lensing (IAU Symp. 173),
ed. C. S. Kochanek, \& J. N. Hewitt (Dordrecht: Reidel), in press
(astro-ph/9508098) 

\reference{A} Schechter, P. L. 1987, in Structure and Dynamics of Elliptical
Galaxies (IAU Symp. 127), ed. T. de Zeeuw (Dordrecht: Reidel), 217

\reference{A} Stark, A. A., et al. 1992, \apjs, 79, 77

\reference{A} Statler, T. S. 1994a, \apj, 425, 458

\reference{A} Statler, T. S. 1994b, \aj, 108, 111

\reference{A} Trinchieri, G., Fabbiano, G., \& Canizares, C., 1986, \apj,
310, 637

\reference{A} Tr\"{u}mper, J. 1983, Adv. Space Res., 2, 241
 
\end{references}
\end{document}